\documentclass[12pt]{article}

\usepackage{graphicx}

\def\rf#1{(\ref{eq:#1})}
\def\lab#1{\label{eq:#1}}

\def\br{\begin{eqnarray}}
\def\er{\end{eqnarray}}
\def\be{\begin{equation}}
\def\ee{\end{equation}}
\def\({\left(}
\def\){\right)}

\def\rlx{\relax\leavevmode}
\def\IR{\rlx\hbox{\rm I\kern-.18em R}}

\def\vt{\vartheta}
\def\u2{\mid u\mid^2}
\newcommand{\sbr}[2]{\left\lbrack\,{#1}\, ,\,{#2}\,\right\rbrack}

\def\IZ{\rlx\hbox{\sf Z\kern-.4em Z}}
\def\IR{\rlx\hbox{\rm I\kern-.18em R}}
\def\IC{\rlx\hbox{\,$\inbar\kern-.3em{\rm C}$}}
\def\one{\hbox{{1}\kern-.25em\hbox{l}}}

\begin{document}

\begin{titlepage}
\vspace*{-1cm}

\vskip 3cm

\vspace{.2in}
\begin{center}
{\large\bf Exact vortex solutions in a $CP^N$ Skyrme-Faddeev type model }
\end{center}

\vspace{.3cm}

\begin{center}
L. A. Ferreira\footnote{e-mail: laf@ifsc.usp.br} and
P. Klimas\footnote{e-mail:klimas.ftg@gmail.com}

\vspace{.3 in}
\small

\par \vskip .2in \noindent
Instituto de F\'\i sica de S\~ao Carlos; IFSC/USP;\\
Universidade de S\~ao Paulo  \\ 
Avenida Trabalhador S\~ao-carlense 400, Caixa Postal 369\\ 
CEP 13560-970, S\~ao Carlos-SP, Brazil\\

\normalsize
\end{center}

\vspace{.5in}

\begin{abstract}

We consider a four dimensional field theory with target space being
$CP^N$ which constitutes a generalization of the usual Skyrme-Faddeev
model defined on $CP^1$. We show that it possesses an integrable
sector presenting an infinite number of local conservation laws, which
are associated to the hidden symmetries of the zero curvature
representation of the theory in loop space. We construct an infinite
class of exact solutions for that integrable submodel where the fields
are meromorphic functions of the combinations  $\(x^1+i\, x^2\)$ and
$\(x^3+x^0\)$ of the Cartesian coordinates of four dimensional Minkowski
space-time. Among those solutions we have static vortices and also
vortices with waves traveling along them with the speed of light. The
energy per unity of length of the vortices show an interesting and
intricate interaction among the vortices and  waves. 

\end{abstract} 
\end{titlepage}

\section{Introduction}
\label{sec:intro}
\setcounter{equation}{0}

The development of non-perturbative methods is of crucial importance for the
study of strong coupling phenomena in Physics, specially in field
theories. It has become clear that  solitons and hidden symmetries
play an special role in that context. In many theories the solitons
are the suitable normal modes to describe the strong coupling
regime. That is true for instance in the two dimensional sine-Gordon
model \cite{coleman} and the four dimensional supersymmetric gauge
theories \cite{duality}, where a
duality exists between the weak and strong coupling sectors where fundamental
particles and solitons exchange roles. In addition,  the
appearance of solitons is associated to a high degree of symmetries and
conservation laws. An important aspect is that in general those are
not symmetries of the Lagrangian or of the equations of motion, and for that
reason are called hidden symmetries. They  appear in special and
deep structures of the theory, and in the case of integrable two
dimensional field theories they are the symmetries of the zero
curvature condition or the Lax-Zakharov-Shabat equation \cite{lax}. Of
course, it 
would be very important to discover the counter part of those
structures in realistic four dimensional field theories like gauge
theories. A proposal \cite{afs} to approach that problem uses the concept flat
connections on loop
spaces to construct the generalization of the zero curvature condition
for theories in dimensions higher than two. Such approach has obtained
success is several models leading to infinite number of conservation
laws and exact solutions \cite{afs-review}.  

In this paper we explore
the ideas of \cite{afs,afs-review} to study a non-linear four dimensional field
theory which is in fact an extension of the Skyrme-Faddeev model on
the target space $CP^N$. We show that such model possesses an
integrable sector that presents an infinite number of conserved
currents which are not associated to any symmetry of the Lagrangian or
equations of motion. They are hidden symmetries of the representation
of the theory in terms of the generalized zero curvature in loop
space. The integrable submodel is obtained by restricting the theory
by some constraints and a relation among the coupling constants. In
addition, we show that such integrable sector possesses an 
infinite class of exact solutions where the $N$ complex fields of the
model are meromorphic functions of the combinations $ x^1+ix^2$
and $x^3+x^0$, of the Cartesian coordinates $x^{\mu}$, $\mu=0,1,2,3$, of four
dimensional Minkowski space-time. Among those solutions we have static
vortices and also vortices with waves traveling along them with the
speed of light. The static vortices are Bogomolny type solutions
and their energy per unity of length come from boundary terms in the
energy functional. They correspond in fact to the Bogomolny solutions of
the $CP^N$ model in two dimensions \cite{wojtekbook,wojtekdin,wojtekused}, but
not to the static non-Bogomolny solutions of that model.  Our results
constitute a generalization to  
$CP^N$ of the those obtained in \cite{vortexlaf} for an extension of
the usual  Skyrme-Faddeev model \cite{sf} defined on $CP^1$ (or equivalently
$S^2$). Vortex type solutions of the usual $CP^1$ Skyrme-Faddeev model were
also considered in \cite{Hietarinta:vortex,hirayama}, and vortex with
waves were considered in supersymmetric gauge theories in
\cite{nitta}.  It is worth mentioning that the Skyrme-Faddeev model on
the coset space $SU(N+1)/U(1)^N$ has been conjectured to correspond to a
low energy effective theory for pure $SU(N+1)$ Yang-Mills theory
\cite{faddeevsun}. In \cite{kondo} it has been argued that for $N\geq 2$,  the
relevant low 
energy degrees of freedom may also be described by the coset space
$SU(N+1)/SU(N)\otimes U(1)$, or $CP^N$, which the authors call the
{\em minimum case}. Therefore, the results of this paper may play some
role in that approach proposed in \cite{kondo}.         

The paper is organized as follows: in section \ref{sec:model} we
define the model exploring the fact that $CP^N$ is the symmetric space
$SU(N+1)/SU(N)\otimes U(1)$. In section \ref{sec:integrable} we present
the integrable sector possessing an infinite number of conservation
laws, and the exact solution are constructed in subsection
\ref{sec:solutions}. The energies per unity of length of the vortex
solutions are calculated in section \ref{sec:energy}. The spectrum is
quite interesting showing an intricate interaction among the vortices
and waves.  The energy of the static vortices comes from 
boundary terms, as it is common in Bogomolny type solutions. The case
of $CP^2$ is discussed in more detail in subsection \ref{sec:su3case}.

\section{The model}
\label{sec:model}
\setcounter{equation}{0}

We shall explore the fact that $CP^N$ is a symmetric space
\cite{helgason}. Indeed, it 
is a coset space $CP^N=SU(N+1)/SU(N)\otimes U(1)$ with the subgroup
$SU(N)\otimes U(1)$ being invariant under the involutive automorphism
($\sigma^2=1$) 
\be
\sigma\(T\)=\Omega\, T\, \Omega^{-1}\; ; \qquad\qquad
\Omega=e^{i\,\pi\,\Lambda}\; ;\qquad\qquad \Lambda=\frac{2\lambda_N\cdot
  H}{\alpha_N^2} 
\lab{sigmadef}
\ee
where $\lambda_N$ is the last fundamental weight of $SU(N+1)$,
i.e. the highest weight of the ${\bar N}$ representation, and
$\alpha_N$ is the simple root of $SU(N+1)$ associated to that
fundamental weight, i.e. $2\lambda_N\cdot \alpha_N/\alpha_N^2=1$. The invariant
subgroup $SU(N)\otimes U(1)$ is generated by the Cartan
subalgebra generators 
$H_i$, $i=1,2,\ldots N$, and the step operators associated to roots not
containing $\alpha_N$ in its expansion in terms of simple
roots. The space $CP^N=SU(N+1)/SU(N)\otimes U(1)$ has a nice
parametrization in terms of the so-called principal variable
\cite{princvar,olive}, defined
by 
\be
X\(g\)\equiv g\,\sigma(g)^{-1}\qquad \qquad \qquad g\in SU(N+1)
\lab{princvar}
\ee
Indeed, if $k\in SU(N)\otimes U(1)$ then $X\(g\,k\)=X\(g\)$, since
$\sigma\(k\)=k$, and so we
have just one matrix $X\(g\)$ for each coset in $SU(N+1)/SU(N)\otimes
U(1)$. 

We now introduce, in $3+1$ dimensions, a field theory with target
space being $CP^N$ and defined by the Lagrangian 
\br
{\cal L}&=& -\frac{M^2}{2}\,{\rm Tr}\(X^{-1}\,\partial_{\mu}X\)^2
+\frac{1}{e^2}\,{\rm
  Tr}\(\sbr{X^{-1}\,\partial_{\mu}X}{X^{-1}\,\partial_{\nu}X}\)^2 +
\frac{\beta}{2}\, \left[{\rm Tr}\(X^{-1}\,\partial_{\mu}X\)^2\right]^2
\nonumber\\
&+&\gamma\, \left[{\rm Tr}\(X^{-1}\,\partial_{\mu}X\;\;
X^{-1}\,\partial_{\nu}X\)\right]^2
\lab{actionx}
\er
The coupling constant $M^2$ has dimension of
mass, and $e^2$, $\beta$ and $\gamma$ are dimensionless coupling
constants. The derivatives $\partial_{\mu}$ are with respect to the
Cartesian coordinates $x^{\mu}$, $\mu = 0,1,2,3$. 
Notice that it consists of a generalization to $CP^N$ of the Skyrme-Faddeev
model \cite{sf}, and we shall refer to it as the $CP^N$ extended
Skyrme-Faddeev  model 
(CPNSF). In the case of $N=1$ it coincides with an extension of the
Skyrme-Faddeev model considered
in \cite{vortexlaf}.  The presence of terms which are quadractic and
quartic in derivatives of the $X$ field imply, according to Derrick's
theorem,  that one can have stable
static solutions in three spatial dimensions. However, we will be
concerned in this paper with exact static and time-dependent vortex
solutions of \rf{actionx}. 
  The theory \rf{actionx} has a global left
$SU(N+1)$ symmetry such that $g\rightarrow 
{\bar g}\,g$, with ${\bar g}\, , \, g \in SU(N+1)$, which implies that
$X\rightarrow {\bar g}\, X\, \sigma\( {\bar g}\)^{-1}$, and so
$X^{-1}\,\partial_{\mu}X \rightarrow \sigma\( {\bar g}\)\,
X^{-1}\,\partial_{\mu}X\, \sigma\( {\bar g}\)^{-1}$. In addition it has
 a right local $SU(N)\otimes U(1)$ symmetry such that $g\rightarrow  g\, k$,
 with $k\in SU(N)\otimes U(1)$ and $g\in SU(N+1)$, which implies that
 $X$ is invariant. That symmetry will play a role when we work with
 the group elements $g$ instead of $X$, since $g^{-1}\partial_{\mu}g
 \rightarrow k^{-1}\,g^{-1}\partial_{\mu}g\, k+
 k^{-1}\,\partial_{\mu} k$.

The Euler-Lagrange equations associated to \rf{actionx} are given by
\be
\partial^{\mu}J_{\mu}=0
\lab{eqmot}
\ee
with
\br
J_{\mu}&\equiv& \left[M^2\,
  \eta_{\mu\nu}-2\,\beta\,{\rm Tr}\(X^{-1}\,\partial_{\rho}X\)^2\,
  \eta_{\mu\nu} 
- 4\,\gamma\,{\rm Tr}\(X^{-1}\,\partial_{\mu}X\;
X^{-1}\,\partial_{\nu}X\)\right]\, X^{-1}\,\partial^{\nu}X
\nonumber\\
&+&\frac{4}{e^2}
\sbr{\sbr{X^{-1}\,\partial_{\mu}X}{X^{-1}\,
\partial_{\nu}X}}{X^{-1}\,\partial^{\nu}X} 
\lab{current}
\er
where $\eta_{\mu\nu}={\rm diag}\(1,-1,-1,-1\)$, is the Minkowski 
metric. In order to obtain  \rf{eqmot} we have used the fact that for
any vector quantity $Y_{\mu}$ we have the identities
$\sbr{Y^{\mu}}{Y_{\mu}}=0$ and
$\sbr{Y_{\mu}}{\sbr{Y_{\nu}}{\sbr{Y^{\mu}}{Y^{\nu}}}}=0$. The current
$J_{\mu}$ lies in the algebra of $SU(N+1)$ and so the number of
independent conserved currents is equal to ${\rm
  dim}\,\left[SU(N+1)\right]=N^2+2\,N$. They are the Noether currents
associated to left global $SU(N+1)$ symmetry of \rf{actionx}  mentioned
above.  

Using \rf{princvar} one gets that 
\be
{X^{-1}\,\partial_{\mu}X}= \sigma(g)\, P_{\mu}\,\sigma(g)^{-1}
\lab{prepmudef}
\ee
with $P_{\mu}$ being the odd part, under $\sigma$, of the Maurer-Cartan
form, i.e.
\be
 P_{\mu}= g^{-1}\,\partial_{\mu}g-\sigma\(g^{-1}\,\partial_{\mu}g\)
\lab{pmudef}
\ee
Therefore, one gets that
\be
J_{\mu}=\sigma(g)\, B_{\mu}\,\sigma(g)^{-1}
\lab{conservedcurr}
\ee
with
\br
B_{\mu}&\equiv& \left[M^2\,
  \eta_{\mu\nu}-2\,\beta\,{\rm Tr}\(P_{\rho}\)^2\,
  \eta_{\mu\nu} 
- 4\,\gamma\,{\rm Tr}\(P_{\mu}\,P_{\nu}\)\right]\, P^{\nu}
+\frac{4}{e^2}
\sbr{\sbr{P_{\mu}}{P_{\nu}}}{P^{\nu}} 
\lab{bmudef}
\er
Since $P_{\mu}$ is odd it turns out that so is $B_{\mu}$, i.e. 
\be
\sigma\(B_{\mu}\)=-B_{\mu}
\ee
Then it follows that the equation of motion \rf{eqmot} can be written
as 
\be
\partial^{\mu}B_{\mu}+\sbr{A^{\mu}}{B_{\mu}}=0 \qquad \qquad\qquad 
{\rm with} \qquad A_{\mu}=g^{-1}\partial_{\mu}g
\lab{eqmotab}
\ee
Therefore, the equations of motion of \rf{actionx} takes the form of
the generalized zero curvature conditions proposed in \cite{afs,afs-review}, in
terms of a vector $B_{\mu}$ and a flat connection $A_{\mu}$. We
discuss below how to use the methods of \cite{afs,afs-review} to
construct an infinite number of conserved currents for a submodel of
\rf{actionx}. It turns out in fact that the equations of motion
\rf{actionx} depend only on the even part of the flat connection
$A_{\mu}$. Indeed, subtracting \rf{eqmotab} from its image under
$\sigma$ one gets that 
\be
\partial^{\mu}B_{\mu}+\sbr{a^{\mu}}{B_{\mu}}=0 \qquad \qquad\qquad 
{\rm with} \qquad a_{\mu}\equiv \frac{\(1+\sigma\)}{2}\, A_{\mu}
\lab{eqmotab2}
\ee
The odd part of $A_{\mu}$ is $P_{\mu}$ and that commutes with
$B_{\mu}$ thanks to the identities mentioned below
\rf{current}. Therefore, the equations of motion \rf{eqmotab2} depend
only on the representation of the subgroup $SU(N)\otimes U(1)$ under
which $B_{\mu}$ transforms, under the action of the (not-flat)
connection $a_{\mu}$. But $B_{\mu}$ belongs to the odd subspace of the
algebra of $SU(N+1)$ and that transforms under the $N(1)+{\bar N}(-1)$
representation of $SU(N)\otimes U(1)$. In order to see that, notice
that the generators of $SU(N+1)$ which are odd under the automorphism
$\sigma$ defined in \rf{sigmadef}, are the step operators $E_{\alpha}$
associated
to roots $\alpha$ that contain the simple root $\alpha_N$ in their expansion in
terms of simple roots. Indeed, we have that
\br
S_{\pm i}\equiv E_{\pm(\alpha_i+\alpha_{i+1}+\ldots \alpha_N)}
\quad\quad\qquad\quad \sigma\(S_{\pm i}\)=-S_{\pm i} \quad\qquad\quad\qquad 
i=1,2,\ldots N
\lab{sidef}
\er
In the $(N+1)$-dimensional defining representation of $SU(N+1)$ those
generators are given by the matrices
\be
\(S_{i}\)_{rs}= \delta_{r\, ,\,i}\,\,\delta_{s\,,\,N+1}
\quad\qquad\quad\qquad  
S_{-i}=S_i^{\dagger} \quad\qquad\quad\qquad r,s=1,2,\ldots N+1
\ee  
A basis for the $N^2$ generators of $SU(N)\otimes U(1)$ can be taken
as $\sbr{S_i}{S_{-j}}$, with $i,j=1,2,\ldots N$, and one can easily check
that 
\br
\sbr{\sbr{S_{i}}{S_{-j}}}{S_{k}}&=&\delta_{ij}\,S_{k}+\delta_{jk}\,S_{i}
\nonumber\\
\sbr{\sbr{S_{i}}{S_{-j}}}{S_{-k}}&=&-\delta_{ij}\,S_{-k}-\delta_{ik}\,S_{-j}
\lab{scommrel}
\er
which establishes that the subspaces generated by $S_i$ and $S_{-i}$
correspond indeed to the $N(1)$ and ${\bar N}(-1)$ representations
respectively  of
$SU(N)\otimes U(1)$. The $U(1)$ generator of $SU(N)\otimes U(1)$ corresponds to
the $\Lambda$ operator defined in \rf{sigmadef}, and in such basis is
given by 
\be
\Lambda = \frac{1}{N+1}\, \sum_{i=1}^N\sbr{S_i}{S_{-i}}
\ee
In addition, such subspaces
are abelian  
\be
\sbr{S_{i}}{S_{j}}=\sbr{S_{-i}}{S_{-j}}=0
\ee
and satisfy
\be
{\rm Tr} \(S_{i}\,S_{-j}\)=\delta_{ij} \qquad \qquad \qquad 
{\rm Tr} \(S_{i}\,S_{j}\)={\rm Tr} \(S_{-i}\,S_{-j}\)=0
\lab{traceform}
\ee
We now introduce coordinates in $CP^N$ by providing a suitable
parametrization of the group elements $g \in SU(N+1)$. We follow the
results of section 8 of \cite{erica}, and introduce complex fields
$u_i$ as 
\be
g=e^{i\,u_i\,S_{i}}\,e^{\varphi\,u_i\,u_j^* \sbr{S_i}{S_{-j}}}\, 
e^{i\,u_i^*\,S_{-i}} \; ;
\quad\quad
\quad\qquad
\varphi\equiv \frac{\log\sqrt{1+u^{\dagger}\cdot u}}{u^{\dagger}\cdot
  u}
\ee
In the $(N+1)$-dimensional defining representation of $SU(N+1)$ we
have that $g$ is given by the matrices
\br
g\equiv \frac{1}{\vt}\,\(\begin{array}{cc}
\Delta&i\,u\\
i\,u^{\dagger}&1
\end{array}\) \qquad\qquad \qquad \quad \vt=\sqrt{1+u^{\dagger}\cdot u}
\lab{gdef}
\er
where $\Delta$ is the hermitian $N\times N$-matrix
\br
\Delta_{ij}=\vt\,\delta_{ij}-\frac{u_i\,u_j^*}{1+\vt}
\quad\quad \quad
\mbox{\rm which satisfies} \quad\quad \quad 
\Delta\cdot u= u\; ; \quad\quad u^{\dagger}\cdot \Delta= u^{\dagger}
\lab{deltadef}
\er
In such representation the group element $\Omega$ introduced in
\rf{sigmadef} can be written as
\br
\Omega= e^{i\,\pi/(N+1)}\,\(\begin{array}{cc}
\one_{N\times N}&0\\
0&-1
\end{array}\)
\er
Therefore, for $g$ given by \rf{gdef}, one has $\sigma\(g\)=g^{-1}$, 
and so the principal variable introduced in \rf{princvar} becomes
\br
X\(g\)= g^2 =  \(\begin{array}{cc}
\one_{N\times N}&0\\
0&-1
\end{array}\) + \frac{2}{\vt^2}\,\(\begin{array}{cc}
-u\otimes u^{\dagger}&i\,u\\
i\,u^{\dagger}&1
\end{array}\)
\er
Therefore, under a left translation $X\rightarrow U\, X\,
\sigma\(U\)^{-1}$, with $U$ being an element of $SU(N)\otimes U(1)$,
i.e.
\br
U= e^{i\,\theta/(N+1)}\,\(\begin{array}{cc}
 {\hat U}&0\\
0&e^{-i\,\theta}
\end{array}\) \qquad {\rm with} \qquad {\hat U}\,{\hat
  U}^{\dagger}=\one\; ; 
\qquad {\rm det}\,{\hat U}=1
\er
we have that $u$ transforms under the defining $N$-dimensional
representation of $SU(N)\otimes U(1)$,
i.e. $u\rightarrow e^{i\,\theta}\, {\hat U}\, u$. 

The odd part of the Maurer-Cartan form introduced in \rf{pmudef}
can be written as $P_{\mu}=P_{\mu}^{(+)}+P_{\mu}^{(-)}$,
where 
\br
 P_{\mu}^{(+)}= \frac{2\,i}{\vt^2}\,\sum_{i=1}^N 
\(\Delta\cdot\partial_{\mu}u\)_i \, S_i 
\qquad\qquad\qquad\qquad
P_{\mu}^{(-)}= \frac{2\,i}{\vt^2}\,\sum_{i=1}^N 
\(\partial_{\mu} u^{\dagger}\cdot\Delta\)_i\, S_{-i}
\lab{ppmdef}
\er
which satisfies $\(P_{\mu}^{(+)}\)^{\dagger}=-P_{\mu}^{(-)}$. 
In addition, the even part of the same one-form, namely the connection
$a_{\mu}$ introduced in \rf{eqmotab2}, becomes
\be
a_{\mu}=\frac{\(1+\sigma\)}{2}g^{-1}\partial_{\mu}g=\sum_{i,j=1}^N 
\frac{\kappa_{\mu}^{ij}}{\vt^2}\, \sbr{S_{i}}{S_{-j}}
\lab{amudef}
\ee 
with
\be
\kappa_{\mu}^{ij}=\frac{1}{2}\,\(u^{\dagger}\cdot
\partial_{\mu}u-\partial_{\mu}u^{\dagger}\cdot u\)\,
\frac{u_i\,u^*_j}{\(1+\vt\)^2}+
\frac{\vt}{1+\vt}\(u_i\,\partial_{\mu}u_j^*-\partial_{\mu}u_i\,u_j^*\)
\lab{kappamudef}
\ee
Using \rf{scommrel} and \rf{ppmdef} one can check that 
\be
\sbr{\sbr{P_{\mu}}{P_{\nu}}}{P^{\nu}}=\tau_{\nu}^{\nu}\, P_{\mu} +
\(\tau_{\mu\nu}-2\,\tau_{\nu\mu}\)\,P^{\nu\(+\)}+
\(\tau_{\nu\mu}-2\,\tau_{\mu\nu}\)\,P^{\nu\(-\)}
\lab{trilpepcomm}
\ee
where we have introduced
\be
\tau_{\mu\nu}\equiv {\rm Tr}\(P_{\mu}^{(+)}\,P_{\nu}^{(-)}\) = 
-\frac{4}{\vt^4}\, \partial_{\nu}u^{\dagger}\cdot
\Delta^2\cdot \partial_{\mu}u
\lab{taudef}
\ee
with $\Delta$ being given in \rf{deltadef}, and so
$\(\Delta^2\)_{ij}=\vt^2\,\delta_{ij}-u_i\,u_j^*$.  Notice that
$\tau_{\mu\nu}^*=\tau_{\nu\mu}$, since $\Delta$ is hermitian. One can
then write the operator $B_{\mu}$ introduced in \rf{bmudef} as
\be
B_{\mu}=B_{\mu}^{(+)}+B_{\mu}^{(-)}\qquad\quad {\rm with}\qquad\quad 
B_{\mu}^{(+)}\equiv C_{\mu\nu}\,P^{\nu(+)}\qquad\qquad
B_{\mu}^{(-)}\equiv P^{\nu(-)}\,C_{\nu\mu}
\ee
and where we have introduced the quantity
\be
C_{\mu\nu}\equiv M^2\,\eta_{\mu\nu}
-\frac{4}{e^2}\left[\(\beta\,e^2-1\)\,\tau_{\rho}^{\rho}\,\eta_{\mu\nu}
+\(\gamma\,e^2-1\)\,\tau_{\mu\nu}+\(\gamma\,e^2+2\)\,\tau_{\nu\mu}\right]
\lab{cmunudef}
\ee
which satisfies $C_{\mu\nu}^*=C_{\nu\mu}$. Therefore, we have 
$\(B_{\mu}^{(+)}\)^{\dagger}=-B_{\mu}^{(-)}$. 

From \rf{scommrel} one observes that $B_{\mu}^{(+)}$ and
$B_{\mu}^{(-)}$ transform under different irreducible representations
of $SU(N)\otimes U(1)$, namely $N(1)$ and ${\bar N}(-1)$. Therefore,
the equation \rf{eqmotab2} split 
into two components, one for each one of those  two irreducible
representations. One can check using \rf{scommrel} and
\rf{ppmdef}-\rf{kappamudef} that
\br
\partial^{\mu}B_{\mu}^{(+)}+\sbr{a^{\mu}}{B_{\mu}^{(+)}}&=&
2\,i\,S_{i}\,\frac{\Delta_{ij}}{\vt^2}\,\left[
\partial^{\mu}\(C_{\mu\nu}\partial^{\nu}u_j\) 
- \frac{1}{\vt^2}\,\(u^*_l\,\delta_{jk}+u^*_k\,\delta_{jl}\)\,
C_{\mu\nu}\,\partial^{\mu}u_l\,\partial^{\nu}u_k\right]
\nonumber
\er
with the equation for $B_{\mu}^{(-)}$ being obtained from that by
complex conjugation. The equations of motion are then given by
\be
\(1+u^{\dagger}\cdot u\)\,\partial^{\mu}\(C_{\mu\nu}\partial^{\nu}u_i\) 
- C_{\mu\nu}\,\left[
\(u^{\dagger}\cdot \partial^{\mu}u\)\, \partial^{\nu} u_i+
\(u^{\dagger}\cdot \partial^{\nu}u\)\, \partial^{\mu} u_i\right]=0
\lab{eqmotcmunu}
\ee
together with their complex conjugates. We then have $2\,N$ equations
of motion corresponding to the $2\,N$ fields $u_i$ and $u^*_i$,
$i=1,2,\ldots N$. Notice that \rf{eqmotcmunu} resembles the $CP^N$
equations of motions, and in fact they reduce to it when
$C_{\mu\nu}\rightarrow \eta_{\mu\nu}$.

\section{The integrable sector}
\label{sec:integrable}
\setcounter{equation}{0}

We now use the concepts of \cite{afs,afs-review} of generalized
integrability to construct an infinite number of conserved currents
for a sector of the theory \rf{actionx}.  The vector $B_{\mu}$
appearing in \rf{eqmotab} and \rf{eqmotab2} lies in the adjoint
representation of $SU(N+1)$ and 
transforms under the $N(1)+{\bar N}(-1)$ representation of
$SU(N)\otimes U(1)$. Therefore, the number of conserved currents one
gets in  \rf{conservedcurr}  is equal to the dimension of the adjoint
of $SU(N+1)$. The basic idea is to look for conditions on the fields
that will make the relations \rf{eqmotab} and  \rf{eqmotab2} still
valid when $B_{\mu}$ lives in higher (possibly infinite)
representations. A detailed account of such procedure is given in
\cite{afs,afs-review}. For the model under consideration the relevant
conditions are given by 
\be
\partial_{\mu}u_i\,\partial^{\mu}u_j=0\qquad\qquad \qquad
\mbox{\rm for any  $i,j=1,2,\ldots N$}
\lab{constraint}
\ee
Therefore the integrable sector is 
selected by those solutions of \rf{eqmotcmunu} which also satisfy
\rf{constraint}.  
That is a generalization to $CP^N$ of the constraints used in the
models with target space being $CP^1$, or equivalently the two
dimensional sphere $S^2$ \cite{afs,afs-review}. Such constraints have
already been considered in \cite{erica} in the context of the pure
$CP^N$ model, leading to an infinite number of conserved currents. A
further study of integrable sectors of pure $CP^N$ model is given in
\cite{joaquin}. 
It then follows from \rf{cmunudef} that, when \rf{constraint} holds
true, one has
\be
C_{\mu\nu}\,\partial^{\mu}u_i\,\partial^{\nu}u_j=0
\lab{cmunuconstraint}
\ee
and the equations of motion \rf{eqmotcmunu} reduce to
\be
\partial^{\mu}\(C_{\mu\nu}\partial^{\nu}u_i\)=0
\lab{reducedeqmot}
\ee
As a consequence of that one can then check that  the currents 
\be
J_{\mu}^{G}\equiv \sum_{i=1}^N\left[\frac{\delta G}{\delta\,
  u_i}\,C_{\mu\nu}\,\partial^{\nu}u_i -
\frac{\delta G}{\delta\,
  u_i^*}\,\partial^{\nu}u_i^*\,C_{\nu\mu}\right]
\lab{currentssubmodel}
\ee
are conserved, i.e. $\partial^{\mu}J_{\mu}^{G}=0$, where $G$ is a
functional of $u_i$ and $u^*_i$, but not of their derivatives. The
conservation of the currents follows directly from
\rf{cmunuconstraint} and  \rf{reducedeqmot}. Notice that such currents
were not obtained as Noether currents associated to symmetries of the
submodel defined by the constraints \rf{constraint}. In fact, it is
not even certain that such submodel possesses a regular
Lagrangian. Such currents are related to hidden symmetries of the
generalized zero curvature based on infinite dimensional
representations of $SU(N+1)$ (see \cite{erica,afs,afs-review} for details).    
 
\subsection{Exact solutions}
\label{sec:solutions}

Notice that due to the constraints \rf{constraint} we have that
$\tau_{\nu\mu} \partial^{\nu}u_i=0$. Therefore the last term in
$C_{\mu\nu}$, given in \rf{cmunudef}, drops out when contracted with
$\partial^{\nu}u_i$. Therefore the reduced equations of motion
\rf{reducedeqmot} can be written as, using \rf{cmunudef} and \rf{taudef},  
\be
M^2\,\partial^2 u_i
+\frac{16}{e^2}\partial^{\mu}\left[\(\partial^{\nu}u^{\dagger}\cdot
  \frac{\Delta^2}{\vt^4}\)_j\, \left[\(\beta\,e^2-1\)
  \, \partial_{\nu}u_j\,\partial_{\mu}u_i+
\(\gamma\, e^2-1\)  \, \partial_{\mu}u_j\,\partial_{\nu}u_i
\right]\right]=0
\lab{reducedeqmot2}
\ee
One can check that 
\be
\partial^{\mu}\(\partial^{\nu}u^{\dagger}\cdot
  \frac{\Delta^2}{\vt^4}\)_j= R^{\mu\nu}_j+N^{\mu\nu}_j
\ee
with
\be
R^{\mu\nu}_j\equiv \left[\vt^2\,\partial^{\mu}\partial^{\nu}u^*_k-
\left[\(\partial^{\mu}u^{\dagger}\cdot u\)\,\partial^{\nu}u^*_k+
\(\partial^{\nu}u^{\dagger}\cdot
u\)\,\partial^{\mu}u^*_k\right]\right]\, \frac{\(\Delta^2\)_{kj}}{\vt^6}
\ee
and
\be
N^{\mu\nu}_j\equiv
-\left[\(\Delta^2\)_{lj}\, u^*_k+\(\Delta^2\)_{lk}\, u^*_j\right]\,
\frac{\partial^{\mu}u_k\,\partial^{\nu}u^*_l}{\vt^6}
\ee
Notice that $R^{\mu\nu}_j=R^{\nu\mu}_j$, and $R_{j,\mu}^{\mu}$ is
proportional to the $CP^N$ equations of motion. In addition, due to
the constraint \rf{constraint} we have that 
$N^{\mu\nu}_j\,\partial_{\mu}u_l=0$. Therefore, \rf{reducedeqmot2}
becomes 
\br
M^2\,\partial^2 u_i
&+&\frac{16}{e^2}\(\partial^{\nu}u^{\dagger}\cdot
  \frac{\Delta^2}{\vt^4}\)_j\, \partial^{\mu}\left[\(\beta\,e^2-1\)
  \, \partial_{\nu}u_j\,\partial_{\mu}u_i+
\(\gamma\, e^2-1\)  \, \partial_{\mu}u_j\,\partial_{\nu}u_i\right]
\nonumber\\
&+&\frac{8}{e^2} \,\(\beta\,e^2+\gamma\, e^2-2\)\,R^{\mu\nu}_j\, 
\( \partial_{\mu}u_j\,\partial_{\nu}u_i+\partial_{\mu}u_i\,\partial_{\nu}u_j\)
=0
\lab{reducedeqmot3}
\er
Let us now introduce the coordinates
\br
z=x^1+i\,\varepsilon_1\,x^2\; ; \quad\;
{\bar z}=x^1-i\,\varepsilon_1\,x^2\; ;
\quad\;
y_{+}=x^3+\varepsilon_2\,x^0\; ; \quad\;
y_{-}=x^3-\varepsilon_2\,x^0
\lab{zydef}
\er
with $\varepsilon_a=\pm 1$, $a=1,2$. Then the metric is 
\be
d\,s^2= \eta_{\mu\nu}\,dx^{\mu}\,dx^{\nu}= - dz\,d{\bar z} -
  dy_{+}\,dy_{-}
\lab{metricholomorphic}
\ee
If we now assume that all the $u_i$'s are functions of $z$ and
$y_{+}$ only, i.e.
\be
u_i=u_i\(z, y_{+}\)\qquad\qquad {\rm and}\qquad\qquad 
u_i^*=u_i^*\({\bar z}, y_{+}\)
\lab{nicesolutions}
\ee
then it follows that 
\be
\partial^2u_i=0 \; ; \qquad\qquad
\partial^{\mu}u_i\,\partial_{\mu}u_j=0  \; ; \qquad\qquad
\partial^{\mu}\left[\partial_{\nu}u_i\,\partial_{\mu}u_j\right]=0
\ee
If in addition we choose the coupling constants such that
\be
\beta\,e^2+\gamma\, e^2-2=0
\lab{couplingcond}
\ee
Then the equations  \rf{reducedeqmot3} as well the constraints
\rf{constraint} are satisfied. We then get an infinite class of exact
solutions for the theory \rf{actionx}, given by the configurations
\rf{nicesolutions}. 

\section{The energy}
\label{sec:energy}
\setcounter{equation}{0}

Using \rf{prepmudef}, \rf{trilpepcomm},  \rf{taudef} and \rf{cmunudef}
one can write the Lagrangian \rf{actionx} as
\be
{\cal L}=
-\frac{1}{2}\,\left[M^2\,\eta_{\mu\nu}+C_{\mu\nu}\right]\,\tau^{\nu\mu}
\lab{actioncmunu}
\ee
Therefore, any variation of ${\cal L}$ w.r.t. the fields leads to
\be
\delta{\cal L}= - C_{\mu\nu}\,\delta \tau^{\nu\mu}
\ee
The Hamiltonian density is given by
\br
{\cal H}&=&\frac{\delta\, {\cal L}}{\delta\,\partial_{0}u_i}\,\partial_0 u_i 
+\frac{\delta\, {\cal L}}{\delta\,\partial_{0}u_i^*}\,\partial_0 u_i^* 
- {\cal L}\nonumber\\
&=&
\frac{1}{2}\,\left[M^2\,\eta_{\mu\nu}+C_{\mu\nu}\right]\,\tau^{\nu\mu} 
- C^{\rho 0}\,\tau_{0\rho} - C^{0\rho}\,\tau_{\rho 0}
\nonumber\\
&=& -M^2\,\(\tau_{00}+\tau_{aa}\) 
+\frac{2}{e^2}\left[
\(\beta\,e^2-1\)\,\tau_{\rho}^{\rho}\,\(3\,\tau_{00}+ \tau_{aa}\) 
+\(\gamma\,e^2-1\)\(4\,\tau^{\rho 0}\,\tau_{0\rho}-\tau^{\rho\nu}\,\tau_{\nu\rho}\)
\right. \nonumber\\
&+&\left. 
\(\gamma\,e^2+2\)\(2\,\tau^{0\rho}\,\tau_{0\rho}+2\,\tau^{\rho 0}\,\tau_{\rho 0}
-\tau^{\rho\nu}\,\tau_{\rho\nu}\)\right]
 \er
where $a=1,2,3$, stands for the space coordinates $x^a$. 
Notice that the last term, proportional to $\(\gamma\,e^2+2\)$,
vanishes when the constraints \rf{constraint} are imposed. In
addition, when the condition \rf{couplingcond}, among the coupling
constants, are taken into account we have that the Hamiltonian reduces
to
\br
{\cal H}_c&=&4\,M^2\,\frac{\(\partial_0u^{\dagger}\cdot\Delta^2\cdot\partial_0u+ 
\partial_au^{\dagger}\cdot\Delta^2\cdot\partial_au\)}{\(1+u^{\dagger}\cdot
u\)^2}
\nonumber\\
&+&16\,\(\beta-\gamma\)\,
\frac{\(\Delta^2\)_{ij}\,\(\Delta^2\)_{kl}}{\(1+u^{\dagger}\cdot u\)^4}\,
\left[\(\partial_0 u_j\,\partial_a u_l-\partial_0
  u_l\,\partial_au_j\)
\(\partial_au^*_i\,\partial_0u^*_k 
- \partial_0u^*_i\,\partial_au^*_k\)
\right. \nonumber\\ 
&+& \left. \sum_{a< b}
\(\partial_au_j\,\partial_bu_l-\partial_au_l\,\partial_bu_j\) 
\(\partial_bu^*_i\,\partial_a u^*_k-\partial_au^*_i\,\partial_b u^*_k\)\right] 
\er 
with $a,b=1,2,3$, and $i,j,k,l=1,2,\ldots N$. Notice that the case
$\beta =\gamma$ is special, since the reduced Hamiltonian becomes positive
definite. 
For the solutions of the type we are considering, namely
\rf{nicesolutions}, we get
\br
{\cal H}_c&=&8\,M^2\,\frac{
\(\partial_{{\bar z}}u^{\dagger}\cdot\Delta^2\cdot\partial_zu+ 
\partial_{y^{+}}u^{\dagger}\cdot\Delta^2\cdot\partial_{y^{+}}u\)}
{\(1+u^{\dagger}\cdot u\)^2}
\lab{energydensityholomorphic}
\\
&+&64\,\(\beta-\gamma\)\,
\frac{\(\Delta^2\)_{ij}\,\(\Delta^2\)_{kl}}{\(1+u^{\dagger}\cdot u\)^4}\,
\left[\(\partial_{y^{+}} u_j\,\partial_zu_l-\partial_{y^{+}}
  u_l\,\partial_zu_j\)
\(\partial_{{\bar z}}u^*_i\,\partial_{y^{+}}u^*_k 
- \partial_{y^{+}}u^*_i\,\partial_{{\bar z}}u^*_k\)
\right] \nonumber
\er 
Let us first consider the static solutions. If a given solution of the
type \rf{nicesolutions} does not depend upon the time, then it does not
depend upon $x^3$ as well. Therefore, the contribution to the energy
comes only from the first term in \rf{energydensityholomorphic}. 
Notice however that for the solutions \rf{nicesolutions} (static or
not) one has the identity 
\be
\partial_z\,\partial_{{\bar z}}\,\ln\(1+u^{\dagger}\cdot u\)= 
\frac{\partial_{{\bar
      z}}u^{\dagger}\cdot\Delta^2\cdot\partial_zu}{\(1+u^{\dagger}\cdot
  u\)^2} 
\lab{niceidentity}
\ee
Therefore, the energy per unit of length for the static solutions is given
by
\br
{\cal E}_{{\rm static}}&=& \int dx^1\,dx^2\, {\cal H}_c = 8\, M^2\,
\int dx^1\,dx^2\,\partial_z\,\partial_{{\bar
    z}}\,\ln\(1+u^{\dagger}\cdot u\)
\lab{staticvortexenergy}
\er
So, the problem reduces to that of the $CP^N$ model in two Euclidean
dimensions, i.e. our static vortex solutions have an energy per unit
of length which equals the Euclidean action of the $CP^N$ lumps. As
shown by \cite{luscher} the finite energy (action) solutions are those where
the $u$ fields are rational functions, i.e.
\be
u_i=\frac{p_i\(z\)}{q_i\(z\)}
\lab{rational}
\ee
where $p_i\(z\)$ and $q_i\(z\)$ are polynomials in the $z$
variable. Following the arguments of \cite{luscher,wojtekbook} (see
also \cite{mantonbook}) one finds that \rf{staticvortexenergy} is
essentially equal to the number of poles of $u_i$'s including those at
infinity, i.e. 
\br
{\cal E}_{{\rm static}}&=& 8\, \pi\, M^2\,\left[ d_{\rm max} +
  \sum_{z_0^{(s)}} h_{\rm max}^{(s)}\right]
\lab{staticenergygeneral}
\er
where $d_{\rm max}$ is the highest degree of the polynomials $p_i$'s,
and the sum is over the zeroes of the polynomials $q_i$'s, with
$h_{\rm max}^{(s)}$ being the highest order of the zeroes of $q_i$'s
at $z=z_0^{(s)}$. Notice therefore that the energy per unit of length
of the static vortex does not depend upon their relative position on
the $x^1\,x^2$ plane, as long as they are all parallel to the
$x^3$-axis. As an example consider the solutions of the form
\be
u_i= c_i\,\(\frac{z}{r_0}\)^{n_i}\,= 
c_i\,\(\frac{\rho}{r_0}\)^{n_i}\, e^{i\,\varepsilon_1\,n_i\,\varphi}
\qquad\qquad i=1,2,\ldots N
\ee
with $c_i$ being complex constants, $n_i$ being integers in order for
the solution to be single 
valued, $r_0$ being a length scale, and where we have introduced
polar coordinates on the 
$x^1\,x^2$ plane, i.e. $z=x^1+ i\,\varepsilon_1\,x^2=\rho\,
e^{i\,\varepsilon_1\,\varphi}$, ($\varepsilon_1=\pm 1$). 
Therefore, for such static vortices we have
\be 
{\cal E}_{{\rm static}}=8\,\pi\,M^2\,\( n_{\rm max} +\mid n_{\rm min} \mid\)
\lab{staticenergy2}
\ee
where $n_{\rm max}$ is the highest positive integer in the set $n_i$,
$i=1,2,\ldots N$ (which corresponds to the highest degree $d_{\rm
  max}$ of the
polynomials $p_i$ in \rf{rational}), and $n_{\rm min}$ is the lowest
negative integer in 
the same set, such that the corresponding complex constants $c_i$ are
non-vanishing. We have in such case zeroes at $z=0$ only, and so
$(-n_{\rm min})$ corresponds to $h_{\rm max}^{(1)}$, with
$z_0^{(1)}=0$. Notice that such result is independent of the number of 
$n_i$'s equal to $n_{\rm max}$ or $n_{\rm min}$. 

Again following the reasoning of \cite{luscher} we can associated 
topological charges to those vortex solutions. The fields $u_i$ provide
a mapping from the $x^1\,x^2$ plane into $CP^N$. However, in order to
have a finite energy per unit length one needs the fields to go to a
constant at infinity on that plane. Therefore, as long as topological
properties are concerned we can consider the $x^1\,x^2$ plane
compactified into the sphere $S^2$. Then the finite energy solutions
define maps 
from $S^2$ to $CP^N$, and they can be classified into the homotopy
classes of $\pi_2\(CP^N\)$. There exists however a theorem
\cite{luscher,goddard,mermin} stating that $\pi_2\(G/H\)=\pi_1\(H\)_G$, where
$\pi_1\(H\)_G$ is the subset of $\pi_1\(H\)$ formed by closed paths in
$H$ which can be contracted to a point in $G$. Since
$CP^N=SU\(N+1\)/SU(N)\otimes U(1)$, the topological charges are given
by $\pi_1\(SU(N)\otimes U(1)\)_{SU(N+1)}$. According to \cite{luscher}
the topological charges of the configurations \rf{rational} are equal
to the number of poles of $u_i$, including those at
infinity. Therefore, the energy per unit length of the vortex
solutions \rf{rational}, given by \rf{staticenergygeneral}, is
proportional to the topological charge, as it is usual in Bogomolny
type solutions. 

We now show that the energy of vortices dependent upon $y_{+}$ is
related to some Noether charges. The Lagrangian \rf{actioncmunu} (or
equivalently \rf{actionx}) is invariant under the phase
transformations $u_i\rightarrow e^{i\,\alpha_i}\, u_i$, $i=1,2,\ldots
N$, with $\alpha_i$ being constant parameters.  Those transformations
correspond in fact to a $U(1)^N$ subgroup of $SU(N)\otimes U(1)$, and
so of $SU(N+1)$. The Noether currents associated to such symmetry are
given by
\br
J_{\mu}^{(i)}=-\frac{4\,i}{\vt^4}\, \sum_{j=1}^N\left[
u_i^*\, \(\Delta^2\)_{ij}\,C_{\mu\nu}\,\partial^{\nu}u_j -
\partial^{\nu}u_j^*\,C_{\nu\mu}\, \(\Delta^2\)_{ji}\,u_i\right]
\lab{u1currents}
\er
Notice that the submodel defined by the constraints \rf{constraint} is
also invariant by those phase transformations, and in fact the
currents \rf{u1currents} can be obtained from \rf{currentssubmodel} by
taking the functions $G$ as $G^{(i)}=u_i\,u_i^*/\(1+u^{\dagger}\cdot
u\)$. 

If one imposes the constraints \rf{constraint} and the condition
\rf{couplingcond} on the coupling constants, then the currents
\rf{u1currents} become (the upper index $c$ stands for constrained currents)
\br
{J^c}_{\mu}^{(i)}&=&-\frac{4\,i}{\vt^4}\, M^2\,\sum_{j=1}^N\left[ 
u_i^*\, \(\Delta^2\)_{ij}\,\partial_{\mu}u_j-
\partial_{\mu}u_j^*\, \(\Delta^2\)_{ji}\,u_i\right]\nonumber\\
&-&\frac{32\,i}{\vt^8}\, \(\beta-\gamma\)\,\sum_{j,k,l=1}^N\left[
\(\Delta^2\)_{ij} \,\(\Delta^2\)_{kl}\,u_i^*\,\partial^{\nu}u_k^*\,
\(\partial_{\mu}u_j\,\partial_{\nu}u_l-\partial_{\nu}u_j\,\partial_{\mu}u_l\)
\right. \nonumber\\
&-&\left.\(\Delta^2\)_{ji} \,\(\Delta^2\)_{lk}\,u_i\,\partial^{\nu}u_k\,
\(\partial_{\mu}u_j^*\,\partial_{\nu}u_l^*-\partial_{\nu}u_j^*\,\partial_{\mu}u_l^*
\)\right]
\er
If one now considers solutions of the class \rf{nicesolutions} of the
form 
\be
u_i= v_i\(z\)\,e^{i\,k_i\,y_{+}}
\ee
with  $k_i$ being the inverse of a wavelength, then the energy density
\rf{energydensityholomorphic} can be written as
\be
{\cal H}_c= 8\, M^2\,
\partial_z\,\partial_{{\bar z}}\,\ln\(1+v^{\dagger}\cdot v\) 
+ \varepsilon_2\,\sum_{i=1}^N k_i\,{J^c}_{0}^{(i)}
\lab{niceenergynoether}
\ee
where we have used \rf{niceidentity}, and $\varepsilon_2$ is defined
in \rf{zydef}. 

Therefore, using the results leading to \rf{staticenergy2}, one
obtains that for vortex solutions of the form
\be
u_i= c_i\,\(\frac{z}{r_0}\)^{n_i}\, e^{i\,k_i\,y_{+}}= 
c_i\,\(\frac{\rho}{r_0}\)^{n_i}\, e^{i\,\varepsilon_1\,n_i\,\varphi}\,
e^{i\,k_i\,\(x^3+\varepsilon_2\,x^0\)}
\lab{nicevortexsol}
\ee
the energy per unit length is given by
\be
{\cal E}_{{\rm vortex/wave}}= \int dx^1\,dx^2\, {\cal H}_c=
8\,\pi\,M^2\,\( n_{\rm max} +\mid n_{\rm min} \mid\)
+ \varepsilon_2\,\sum_{i=1}^N k_i\,Q^{(i)}
\lab{vortexwaveenergy}
\ee
where $Q^{(i)}$ are the Noether charges per unit length associated to the phase
transformations $u_i\rightarrow e^{i\alpha_i}\,u_i$, i.e. 
\br
Q^{(i)}=\int dx^1\,dx^2\,{J^c}_{0}^{(i)}
\er
For the solutions of the type \rf{nicevortexsol} one obtains that
\br
Q^{(i)}&=&8\,\pi\,M^2\,\varepsilon_2\, r_0^2\;
\left[k_i\,\mid c_i\mid^2\,{\cal I}_{(n_i,2,\vec{n},\vec{c})}
+\sum_{j=1}^N(k_i-k_j)\,
\mid c_i\mid^2\,\mid c_j\mid^2\,{\cal I}_{(n_i+n_j,2,\vec{n},\vec{c})}
\right]\nonumber\\  
&-&128\pi\,(\beta-\gamma)\,\varepsilon_2\,
\left[\sum_{j=1}^Nn_j\,(k_i\,n_j-k_j\,n_i)\,
\mid c_i\mid^2\,\mid c_j\mid^2{\cal I}_{(n_i+n_j-1,2,\vec{n},\vec{c})}
\right. \nonumber\\
&-&\left.
\sum_{j,k=1}^N
n_k\,\left[k_i\,n_j-k_j\,n_i+k_j\,n_k-k_k\,n_j\right]\,
\mid c_i\mid^2\,\mid c_j\mid^2\, \mid c_k\mid^2\,
{\cal I}_{(n_i+n_j+n_k-1,3,\vec{n},\vec{c})}\right]\nonumber 
\er
where we have introduced the integrals 
\be
{\cal I}_{\(a,b,{\vec n},{\vec c}\)}\equiv \int_0^{\infty} d\zeta\, 
\frac{\zeta^a}{\left[1+\sum_{k=1}^N\mid c_k\mid^2\,\zeta^{n_k}\right]^b}
\lab{integrals}
\ee
where the integration variable is given by $\zeta\equiv
\rho^2/r_0^2$, and ${\vec n}$ and ${\vec c}$ stand for the set of
integers $n_i$ and constants $c_i$ respectively, 
i.e. ${\vec n}=\(n_1,n_2,\ldots n_N\)$, and ${\vec
  c}=\(c_1,c_2,\ldots c_N\)$. 
 
Therefore the second term in \rf{vortexwaveenergy} becomes
\br
\varepsilon_2\sum_{i=1}^N k_iQ^{(i)}&=&
8\pi M^2 r_0^2
\left[\sum_{i=1}^N k_i^2\mid c_i\mid^2{\cal I}_{(n_i,2,\vec{n},\vec{c})}
+\sum_{\frac{i,j=1}{i<j}}^N(k_i-k_j)^2
\mid c_i\mid^2\mid c_j\mid^2{\cal I}_{(n_i+n_j,2,\vec{n},\vec{c})}
\right]\nonumber\\  
&-&64\pi\,(\beta-\gamma)\,
\left[\sum_{i,j=1}^N(k_i\,n_j-k_j\,n_i)^2\,
\mid c_i\mid^2\,\mid c_j\mid^2{\cal I}_{(n_i+n_j-1,2,\vec{n},\vec{c})}
\right. 
\lab{pieceofenergy}\\
&-&\left.
2\sum_{i,j,k=1}^N\(k_i\,n_j-k_j\,n_i\)\(k_i\,n_k-k_k\,n_i\)
\mid c_i\mid^2\mid c_j\mid^2 \mid c_k\mid^2
{\cal I}_{(n_i+n_j+n_k-1,3,\vec{n},\vec{c})}\right]\nonumber 
\er
Notice that in the double  sums in \rf{pieceofenergy},
whenever the two indices are equal the cor\-res\-pon\-ding coefficients
vanish. In the last 
term, involving a triple sum, the coefficients vanish whenever the
indices in two
pairs are equal (namely $\(i,j\)$ and $\(i,k\)$) but not
when the indices in the third pair are equal (namely
$\(j,k\)$). Therefore, following the analysis of the appendix
\ref{sec:appendix} we conclude that the contribution to the energy per
unity of length, given by \rf{pieceofenergy}, is finite if
\be
 2\, n_{\rm max}>1+n_i \quad {\rm and}\quad      2\, n_{\rm
   max}>1+n_i+n_j \;\; (i\neq j) \qquad
{\rm when}\quad \beta = \gamma
\ee
where $n_{\rm max}$ is the highest positive integer in the set 
${\vec n}=\(n_1, \ldots n_N\)$, such that the corresponding constant
$c_i$ is non-vanishing. Now, if $\beta \neq \gamma$ we need in
addition  the following conditions 
\be
3\, n_{\rm max}>n_i+n_j+n_k\; ; \qquad 2 \mid n_{\rm min}\mid
>-n_i-n_j\;  ;
\qquad 3\mid n_{\rm min}\mid >-n_i-n_j-n_k
\ee
with $i\neq j$ and $i\neq k$, and where $n_{\rm min}$ is the lowest
negative integer in the set ${\vec n}=\(n_1, \ldots n_N\)$, such that
the corresponding constant $c_i$ is non-vanishing. 

Let us make some comments about the structure of the energy per unit
length of the vortices as given by \rf{vortexwaveenergy} and
\rf{pieceofenergy}. Consider the case where all the integers $n_i$ and
wave vectors $k_i$ are equal, i.e. $n_i\equiv n$ and $k_i\equiv k$,
for $i=1,2,\ldots N$. Then \rf{nicevortexsol} becomes 
\be
{\vec u}= {\vec c}\,\(\frac{z}{r_0}\)^{n}\, e^{i\,k\,y_{+}}
\lab{nicevortexsolsu2}
\ee
Therefore we have that $n_{\rm max} = n$, if $n>0$, or $n_{\rm min}=n$ if
$n<0$. In addition, all the terms in \rf{pieceofenergy} vanish except
for the first one. We have in fact a $CP^1$ vortex pointing in a given
fixed direction in $CP^N$, and the energy density
\rf{vortexwaveenergy} reduces to the case $CP^1$ discussed in
\cite{vortexlaf}. Indeed, we have that \rf{vortexwaveenergy} becomes 
\be
{\cal E}_{{\rm vortex/wave}}=8\pi M^2\left[\mid n\mid+ 
\frac{ r_0^2\,k^2}{\mid
  {\vec c}\mid^{2/n}} \, \frac{1}{\mid n\mid}\,
\Gamma\(\frac{\mid n\mid+1}{\mid n\mid}\)\,
\Gamma\(\frac{\mid n\mid-1}{\mid n\mid}\) \right]
\lab{energycp1}
\ee
where $\mid {\vec c}\mid^2\equiv \sum_{i=1}^N \mid c_i\mid^2$, and
where we have rescaled $\zeta\rightarrow \zeta/\mid {\vec
  c}\mid^{2/n}$, and used the fact that
$\int_0^{\infty}d\zeta\frac{\zeta^n}{\(1+\zeta^n\)^2}= 
\int_0^{\infty}d\zeta\frac{\zeta^{-n}}{\(1+\zeta^{-n}\)^2}=
\frac{1}{\mid n\mid}\,
\Gamma\(\frac{\mid n\mid+1}{\mid n\mid}\)\,
\Gamma\(\frac{\mid n\mid-1}{\mid n\mid}\)$. Notice that an equivalent
result would have been obtained by setting 
all the $c_i$'s to zero except for one of them. The second term in
\rf{energycp1} is the energy coming from the coupling of the
wave with the vortex. It grows with $k^2$ which accounts for the
kinetic energy of the wave. However, as $\mid n\mid \rightarrow
\infty$ that term behaves as $1/\mid n\mid$, and so the interaction
between the wave and vortex decreases as $\mid n\mid$ increases. In
addition, we notice that the factor involving $\mid {\vec c}\mid$
depends on the sign of $n$. Therefore, if $n>0$ we see that the energy
from the interaction  between wave and vortex decreases with the increase
of  $\mid {\vec c}\mid$, and that behavior reverts if $n<0$.  That
result is related to the fact that the energy depends upon the length
scale $r_0$, and from \rf{nicevortexsolsu2} we see that $\mid {\vec
  c}\mid$ rescales $r_0$ differently for different signs of $n$.   

In fact the scaling effects of ${\vec c}$ on the energy density
\rf{vortexwaveenergy} can be inferred by considering the case where $c_i =
\lambda^{n_i}\, e^{i\,\theta_i}$, with $\lambda$ real and
positive. From \rf{pieceofenergy} and \rf{integrals} one observes that
$\lambda$ can be absorbed into $\zeta$ by the rescaling $\zeta
\rightarrow \lambda^2 \,\zeta$. Then all the factors $\mid c_i\mid^2$
disappear from  \rf{pieceofenergy} and \rf{integrals}, and everything
is rescaled by $1/\lambda^2$ due to the measure of the integrals
$d\zeta$. So one gets that $\varepsilon_2\sum_{i=1}^N
k_iQ^{(i)}\rightarrow \frac{1}{\lambda^2}\,\varepsilon_2\sum_{i=1}^N
k_iQ^{(i)}$. Therefore, the energy density coming from the interaction
between wave and vortex (second term in \rf{vortexwaveenergy}),
decreases as $\lambda$ increases. At the same time, as $\lambda$
increase we have that $c_i$ increases for $n_i$ positive and decrease
for $n_i$ negative.  

Notice that the term proportional to $\(\beta-\gamma\)$ in
\rf{pieceofenergy} drops if the vectors ${\vec n}$ and ${\vec k}$ are
proportional, i.e. $k_i=k\, n_i$. That implies that we have $u_i =
c_i\, v^{n_i}$  with $v= \frac{z}{r_0}\, e^{i\,k\,y_{+}}$.

\subsection{The case $N=2$}
\label{sec:su3case}

In the case $N=2$ we have that the expression \rf{pieceofenergy}
becomes
\br
\varepsilon_2\sum_{i=1}^2 k_iQ^{(i)}&=&
8\pi M^2 r_0^2
\left[ k_1^2\mid c_1\mid^2{\cal I}_{(n_1,2,\vec{n},\vec{c})}
+k_2^2\mid c_2\mid^2{\cal I}_{(n_2,2,\vec{n},\vec{c})}
\right. \nonumber\\
&+& \left. (k_1-k_2)^2
\mid c_1\mid^2\mid c_2\mid^2{\cal I}_{(n_1+n_2,2,\vec{n},\vec{c})}
\right]\nonumber\\  
&-&128\pi\,(\beta-\gamma)\,\mid c_1\mid^2\mid c_2\mid^2\, 
(k_1\,n_2-k_2\,n_1)^2\,
\left[{\cal I}_{(n_1+n_2-1,2,\vec{n},\vec{c})}
\right. 
\lab{pieceofenergy2}\\
&-&\left.
\mid c_1\mid^2\,{\cal I}_{(2\,n_1+n_2-1,3,\vec{n},\vec{c})}
-\mid c_2\mid^2\,{\cal I}_{(2\,n_2+n_1-1,3,\vec{n},\vec{c})}
\right]\nonumber 
\er

We show below the lowest non-divergent energies per unity length, as
given in \rf{vortexwaveenergy}, for
the vortices with waves traveling along them, and taking $\mid
c_1\mid^2=\mid c_2\mid^2=1$. 

\begin{tabular}{|c|l|}
\hline
$\(n_1,n_2\)$ & ${\cal E}_{{\rm vortex/wave}}/8\,\pi\,M^2$\\
\hline
\hline
$(2,0)$&
$2+\frac{\sqrt{2}}{16}\pi  r_0^2 \left(4 k_1^2-4 k_1
   k_2+3 k_2^2\right)-\frac{4}{M^2} k_2^2
   (\beta -\gamma )$\\
\hline
$(3,0)$&
$3+\frac{2 \sqrt[3]{2}}{9 \sqrt{3}} \pi  r_0^2
   \left(k_1^2-k_1 k_2+k_2^2\right)-\frac{6}{M^2} k_2^2 (\beta -\gamma )$\\
\hline
$(2,-1)$&
$3+ r_0^2 \left(0.83 k_1^2-0.48
   k_1 k_2+0.44 k_2^2\right)-\frac{5.89}{8 M^2}
   (\beta - \gamma ) (k_1+2 k_2)^2$\\
\hline
$(3,1)$&
$3+ r_0^2 \left(0.83 k_1^2-1.18
   k_1 k_2+0.79 k_2^2\right)-\frac{5.89}{8 M^2}
   (\beta - \gamma ) (k_1-3 k_2)^2$\\
\hline
$(-3,-2)$&
$3+ r_0^2 \left(0.44 k_1^2-0.41 k_1
    k_2+0.79 k_2^2\right)-\frac{5.89}{8 M^2} (\beta -\gamma ) (2
    k_1-3 k_2)^2$\\
\hline
$(4,0)$&
$4+\frac{1}{32 \sqrt[4]{2}}\pi  r_0^2 \left(4 k_1^2-4 k_1
   k_2+5 k_2^2\right)-\frac{8}{ M^2} k_2^2
   (\beta -\gamma )$\\
\hline
$(2,-2)$&
$4+\frac{2}{27} \left[\frac{3 \sqrt{3}}{8} \pi r_0^2 \left(5
   k_1^2-2 k_1 k_2+2 k_2^2\right)-\frac{16}{ M^2}
   \left(2 \sqrt{3} \pi -9\right) (\beta -\gamma )
   (k_1+k_2)^2\right]$\\
\hline
$(3,-1)$&
$4+ r_0^2 \left(0.42 k_1^2-0.32
   k_1 k_2+0.35 k_2^2\right)-\frac{4.36}{8 M^2}
   (\beta - \gamma ) (k_1+3 k_2)^2$\\
\hline
$(4,1)$&
$4+ r_0^2 \left(0.42 k_1^2-0.52 k_1
    k_2+0.46 k_2^2\right)-\frac{4.36}{8 M^2} (\beta - \gamma )
    (k_1-4 k_2)^2$\\
\hline
$(5,0)$&
$5+ \frac{r_0^2}{8\pi}
    \left(6.17 k_1^2-6.17 k_1 k_2+9.26
    k_2^2\right)-\frac{10}{ M^2}\( \beta- \gamma\)  k_2^2$\\
\hline
$(-1,4)$&
$5+ r_0^2 \left(0.31 k_1^2-0.24 k_1
    k_2+0.29 k_2^2\right)-\frac{3.45}{8 M^2} (\beta -\gamma ) (4
    k_1+k_2)^2$\\
\hline
$(3,-2)$&
$5+ r_0^2 \left(0.38 k_1^2-0.23 k_1
    k_2+0.26 k_2^2\right)-\frac{3.56}{8 M^2} (\beta - \gamma ) (2
    k_1+3 k_2)^2$\\
\hline
$(2,-3)$&
$5+ r_0^2 \left(0.73 k_1^2-0.22 k_1
    k_2+0.23 k_2^2\right)-\frac{3.56}{8 M^2} (\beta - \gamma ) (3
    k_1+2 k_2)^2$\\
\hline
\end{tabular}

\vspace{2cm}

\noindent{\bf Acknowledgments}

The authors are indebted to Prof. Wojtek Zakrzewski for many useful
discussions on the $CP^N$ solutions. PK is supported by a FAPESP
pos-doc scholarship and LAF is partially supported by CNPq. 

\newpage

\appendix

\section{Analysis of the integrals \rf{integrals}}
\label{sec:appendix}
\setcounter{equation}{0}

Consider the integrals \rf{integrals} and let $n_{\rm max}$ and
$n_{\rm min}$ be the
highest positive and lowest negative integers respectively, in the set
${\vec n}=\(n_1,\ldots n_N\)$, such that the corresponding constants
$c_i$'s are non-vanishing. We then factor out $\zeta^{\mid n_{\rm
    min}\mid}$, and rewrite \rf{integrals} as
\be
{\cal I}_{\(a,b,{\vec n},{\vec c}\)}\equiv \int_0^{\infty} d\zeta\, 
\frac{\zeta^{a+b\,\mid n_{\rm min}\mid}}
{\left[\zeta^{\mid n_{\rm min}\mid}
+\sum_{k=1}^N\mid c_k\mid^2\,\zeta^{n_k+\mid n_{\rm min}\mid}\right]^b}
\lab{integrals2}
\ee
Then all the powers of $\zeta$ in the denominator are non-negative. We
are interested in cases where $b$ equals $2$ or $3$, which are the
ones appearing in \rf{pieceofenergy}. Therefore, for $\zeta
\rightarrow \infty$, we have that the integrand in \rf{integrals2}
behaves as $\zeta^{a-b\,n_{\rm max}}$, and so in order for the
integral to converge we need
\be
b\,n_{\rm max}-a >1
\ee
On the other hand, for $\zeta \rightarrow 0$, the integrand in
\rf{integrals2} behaves as $\zeta^{a+b\,\mid n_{\rm min}\mid}$. Therefore, in
order for the integral to converge we need
\be
 a+b\,\mid n_{\rm min}\mid >-1
\ee
Let us now consider the integrals appearing in the expression
\rf{pieceofenergy}. According to the above analysis the integrals
${\cal I}_{\(n_i,2,{\vec n},{\vec c}\)}$ converges if $2 n_{\rm
  max}>1+n_i$ and if $2\mid n_{\rm min}\mid>-1-n_i$. Notice that
the second condition is always satisfied since, if $n_{\rm min}=0$ we
have that all $n_i$'s are non-negative, and if $n_{\rm min}\neq 0$
then the worst situation happens when $n_i=-\mid n_{\rm min}\mid$, and
the inequality is still satisfied. Therefore
\be
{\cal I}_{\(n_i,2,{\vec n},{\vec c}\)} \qquad\qquad \mbox{\rm converges if}
\qquad \quad 2\, n_{\rm max}>1+n_i
\ee
Now, the integrals ${\cal I}_{\(n_i+n_j,2,{\vec n},{\vec c}\)}$
converges if $2 n_{\rm max}>1+n_i+n_j$ and if $2\mid n_{\rm
  min}\mid>-1-n_i-n_j$. Again the second inequality is always
satisfied since, if $n_{\rm min}=0$ we
have that all $n_i$'s are non-negative, and if $n_{\rm min}\neq 0$
then the worst situation happens when $n_i=n_j=-\mid n_{\rm min}\mid$,
and the condition is still satisfied. 
Therefore
\be
{\cal I}_{\(n_i+n_j,2,{\vec n},{\vec c}\)} \qquad\qquad \mbox{\rm converges if}
\qquad \quad 2\, n_{\rm max}>1+n_i+n_j 
\ee
According to the analysis above we have in addition the following
results
\be
{\cal I}_{\(n_i+n_j-1,2,{\vec n},{\vec c}\)} \qquad \mbox{\rm
  converges if}  \quad 2\, n_{\rm max}>n_i+n_j \; \;{\rm and}\; \;
2\mid n_{\rm min}\mid > - n_i-n_j
\ee
and
\be
{\cal I}_{\(n_i+n_j+n_k-1,3,{\vec n},{\vec c}\)} \quad \mbox{\rm
  converges if}  \quad 3 n_{\rm max}>n_i+n_j+n_k \; \;{\rm and}\; \;
3\mid n_{\rm min}\mid > - n_i-n_j-n_k
\ee


\newpage

\begin{thebibliography}{99}
\bibitem{coleman}
S.~R.~Coleman,
  ``Quantum sine-Gordon equation as the massive Thirring model,''
  Phys.\ Rev.\  D {\bf 11}, 2088 (1975).\\
S.~Mandelstam,
  ``Soliton operators for the quantized sine-Gordon equation,''
  Phys.\ Rev.\  D {\bf 11}, 3026 (1975).

\bibitem{duality}
C.~Montonen and D.~I.~Olive,
  ``Magnetic Monopoles As Gauge Particles?,''
  Phys.\ Lett.\  B {\bf 72}, 117 (1977).\\
C.~Vafa and E.~Witten,
  ``A Strong coupling test of S duality,''
  Nucl.\ Phys.\  B {\bf 431}, 3 (1994)
  [arXiv:hep-th/9408074].\\
N.~Seiberg and E.~Witten,
  ``Electric - magnetic duality, monopole condensation, and confinement in N=2
  supersymmetric Yang-Mills theory,''
  Nucl.\ Phys.\  B {\bf 426}, 19 (1994)
  [Erratum-ibid.\  B {\bf 430}, 485 (1994)]
  [arXiv:hep-th/9407087].

\bibitem{lax}
 P.~D.~Lax,
  ``Integrals Of Nonlinear Equations Of Evolution And Solitary Waves,''
  Commun.\ Pure Appl.\ Math.\  {\bf 21}, 467 (1968).
V.E. Zakharov and A.B. Shabat, {\it Zh. Exp. Teor. Fiz. } {\bf 61} 
(1971) 118-134;  english transl. {\it Soviet Phys. JETP} {\bf 34} (1972) 62-69.

\bibitem{afs}
 O. Alvarez, L.A. Ferreira, J. Sanchez Guillen, ``A
  new approach to 
integrable theories in any dimension'', 
{\em Nucl. Phys.} {\bf B529} (1998) 689-736, [arXiv:hep-th/9710147]

\bibitem{afs-review}
  O.~Alvarez, L.~A.~Ferreira and J.~Sanchez-Guillen,
  ``Integrable theories and loop spaces: fundamentals, applications and new
  developments,''
  Int.\ J.\ Mod.\ Phys.\  A {\bf 24}, 1825 (2009)
  [arXiv:0901.1654 [hep-th]].

\bibitem{wojtekbook}
W.J. Zakrzewski, {\em Low Dimensional Sigma Models\/} (Hilger, Bristol, 1989).

\bibitem{wojtekdin}
  A.~M.~Din and W.~J.~Zakrzewski,
  Nucl.\ Phys.\  B {\bf 174}, 397 (1980).

\bibitem{wojtekused}
  A.~M.~Grundland and W.~J.~Zakrzewski,
``On $CP^1$ and $CP^2$ maps and Weierstrass representations for
surface immersed into multi-dimensional Euclidean spaces'';
{\it Journal of Nonlinear Mathematical Physics} {bf 10}, Number 1,
(2003), 110-135. 

\bibitem{vortexlaf} 
  L.~A.~Ferreira,
  ``Exact vortex solutions in an extended Skyrme-Faddeev model,''
{\em Journal of High Energy Physics} {\bf JHEP05(2009)001}, 
  arXiv:0809.4303 [hep-th].

\bibitem{sf}
  L.~D.~Faddeev, 
  ``Quantization of solitons", Princeton preprint IAS Print-75-QS70
  (1975).\\
  L.~D.~Faddeev, in {it 40 Years in Mathematical Physics}, (World
  Scientific, 1995). \\
 L.~D.~Faddeev and A.~J.~Niemi,
  ``Knots and particles,''
  Nature {\bf 387}, 58 (1997)
  [arXiv:hep-th/9610193].
\\
 P.~Sutcliffe,
  ``Knots in the Skyrme-Faddeev model,''
  Proc.\ Roy.\ Soc.\ Lond.\  A {\bf 463}, 3001 (2007)
  [arXiv:0705.1468 [hep-th]].
\\
  J.~Hietarinta and P.~Salo,
  ``Faddeev-Hopf knots: Dynamics of linked un-knots,''
  Phys.\ Lett.\  B {\bf 451}, 60 (1999)
  [arXiv:hep-th/9811053].
\\
  J.~Hietarinta and P.~Salo,
  ``Ground state in the Faddeev-Skyrme model,''
  Phys.\ Rev.\  D {\bf 62}, 081701 (2000).

\bibitem{Hietarinta:vortex}
J.~Hietarinta, J.~Jaykka and P.~Salo,
``Dynamics of vortices and knots in Faddeev's model'', in {\em
  Workshop on Integrable Theories, Solitons and Duality (2002)},
Proceedings of Science PoS(unesp2002)017,
http://pos.sissa.it/cgi-bin/reader/conf.cgi?confid=8 \\ 
  J.~Hietarinta, J.~Jaykka and P.~Salo,
  ``Relaxation of twisted vortices in the Faddeev-Skyrme model,''
  Phys.\ Lett.\  A {\bf 321} (2004) 324
  [arXiv:cond-mat/0309499].\\
J.~Jaykka and J.~Hietarinta,
  ``Unwinding in Hopfion vortex bunches,''
  arXiv:0904.1305 [hep-th].

\bibitem{hirayama}
  M.~Hirayama, C.~G.~Shi and J.~Yamashita,
  ``Elliptic solutions of the Skyrme model,''
  Phys.\ Rev.\  D {\bf 67}, 105009 (2003)
  [arXiv:hep-th/0303092];\\
 M.~Hirayama and C.~G.~Shi,
  ``A class of exact solutions of the Faddeev model,''
  Phys.\ Rev.\  D {\bf 69}, 045001 (2004)
  [arXiv:hep-th/0310042], \\
 C.~G.~Shi and M.~Hirayama,
  ``Approximate vortex solution of Faddeev model,''
  Int.\ J.\ Mod.\ Phys.\  A {\bf 23}, 1361 (2008)
  [arXiv:0712.4330 [hep-th]].

\bibitem{nitta} M.~Eto, Y.~Isozumi, M.~Nitta and K.~Ohashi,
  ``1/2, 1/4 and 1/8 BPS equations in SUSY Yang-Mills-Higgs systems: Field
  theoretical brane configurations,''
  Nucl.\ Phys.\  B {\bf 752}, 140 (2006)
  [arXiv:hep-th/0506257].
M.~Eto, Y.~Isozumi, M.~Nitta, K.~Ohashi and N.~Sakai,
  ``Solitons in the Higgs phase: The moduli matrix approach,''
  J.\ Phys.\ A  {\bf 39}, R315 (2006)
  [arXiv:hep-th/0602170].

\bibitem{faddeevsun}
  L.~D.~Faddeev and A.~J.~Niemi,
  ``Partial duality in SU(N) Yang-Mills theory,''
  Phys.\ Lett.\  B {\bf 449}, 214 (1999)
  [arXiv:hep-th/9812090].
\\
L.~D.~Faddeev and A.~J.~Niemi,
  ``Decomposing the Yang-Mills field,''
  Phys.\ Lett.\  B {\bf 464}, 90 (1999)
  [arXiv:hep-th/9907180].

\bibitem{kondo}
  K.~I.~Kondo, T.~Shinohara and T.~Murakami,
  ``Reformulating SU(N) Yang-Mills theory based on change of variables,''
  Prog.\ Theor.\ Phys.\  {\bf 120}, 1 (2008)
  [arXiv:0803.0176 [hep-th]].

\bibitem{helgason}
S. Helgason, {\em Differential geometry, Lie groups and symmetric
  spaces}, New York: Academic Press 1978.

\bibitem{princvar}
H.~Eichenherr and M.~Forger,
  ``More About Nonlinear Sigma Models On Symmetric Spaces,''
  Nucl.\ Phys.\  B {\bf 164}, 528 (1980)
  [Erratum-ibid.\  B {\bf 282}, 745 (1987)].

\bibitem{olive}
L.~A.~Ferreira and D.~I.~Olive,
  ``Noncompact Symmetric Spaces And The Toda Molecule Equations,''
  Commun.\ Math.\ Phys.\  {\bf 99}, 365 (1985).

\bibitem{erica}
  L.~A.~Ferreira and E.~E.~Leite,
  ``Integrable theories in any dimension and homogeneous spaces,''
  Nucl.\ Phys.\  B {\bf 547}, 471 (1999)
  [arXiv:hep-th/9810067].

\bibitem{joaquin}
  C.~Adam, J.~Sanchez-Guillen and A.~Wereszczynski,
  ``New integrable sectors in Skyrme and 4-dimensional CP(n) model,''
  J.\ Phys.\ A  {\bf 40}, 1907 (2007)
  [arXiv:hep-th/0610024].

\bibitem{luscher}
A.~D'Adda, M.~Luscher and P.~Di Vecchia,
  ``A 1/N Expandable Series Of Nonlinear Sigma Models With Instantons,''
  Nucl.\ Phys.\  B {\bf 146}, 63 (1978).

\bibitem{mantonbook}
N.~S.~Manton and P.~Sutcliffe,
  ``Topological solitons,''
{\it  Cambridge, UK: Univ. Pr. (2004) 493 p}

\bibitem{goddard}
  P.~Goddard and D.~I.~Olive,
  ``New Developments In The Theory Of Magnetic Monopoles,''
  Rept.\ Prog.\ Phys.\  {\bf 41}, 1357 (1978).

\bibitem{mermin}
  N.~D.~Mermin,
  ``The topological theory of defects in ordered media,''
  Rev.\ Mod.\ Phys.\  {\bf 51}, 591 (1979).

\end{thebibliography}
\end{document}